\documentclass[aps,superscriptaddress,nofootinbib,preprintnumbers]{revtex4}
\usepackage{graphicx,epsfig}
\usepackage{amsmath}
\usepackage{amssymb}
\usepackage{dcolumn}
\usepackage{bm}
\usepackage{amsfonts}
\usepackage{latexsym}

\newcommand{\fr}{\frac}

\newcommand{\lb}{\label}
\newcommand{\be}{\begin{equation}}
\newcommand{\ee}{\end{equation}}
\newcommand{\ba}{\begin{align}}
\newcommand{\ea}{\end{align}}
\newcommand{\bea}{\begin{eqnarray}}
\newcommand{\eea}{\end{eqnarray}}
\newcommand{\bw}{\begin{widetext}}
\newcommand{\ew}{\end{widetext}}

\newcommand{\e}{{\rm e}}

\newcommand{\nn}{\nonumber}
\newcommand{\zt}{\dot{z}}

\newcommand{\y}{\mathbf{y}}

\newcommand{\mm}{\mathrm{m}}
\newcommand{\n}{\mathbf{n}}

\begin{document}

\title{ \begin{flushright}\begin{small}  LAPTH-1344/09\\
CCTP-2010-18
\end{small} \end{flushright} 
Transplanckian bremsstrahlung and black hole production}
\author{Dmitry V. Gal'tsov}\email{galtsov@physics.msu.ru}
\affiliation{Department of Theoretical Physics, Moscow State
University, 119899, Moscow, Russia} \affiliation{Laboratoire de
Physique Th\'eorique LAPTH (CNRS, Universit\'e de Savoie),\\
B.P.110, F-74941 Annecy-le-Vieux cedex, France}

\author{Georgios Kofinas}\email{gkofin@phys.uoa.gr}
\affiliation{Department of Physics and Institute of Theoretical and
Computational Physics, University of Crete, 71003 Heraklion,
Greece}

\author{Pavel Spirin}\email{salotop@list.ru}
\affiliation{Department of Theoretical Physics, Moscow State
University, 119899, Moscow, Russia}
\affiliation{Bogolubov Laboratory of Theoretical Physics, JINR, Joliot-Curie 6, Dubna, Russia}

\author{Theodore N. Tomaras}\email{tomaras@physics.uoc.gr}
\affiliation{Department of Physics and Institute of Theoretical and
Computational Physics,  University of Crete, 71003 Heraklion,
Greece}





\begin{abstract}
Classical gravitational bremsstrahlung in particle collisions at
transplanckian energies is studied in ${\mathcal M}_4\times
{\mathcal T}^d$. The radiation efficiency $\epsilon\equiv E_{\rm
rad}/E_{\rm initial}$ is computed in terms of the Schwarzschild
radius $r_S(\sqrt{s})$, the impact parameter $b$ and the Lorentz
factor $\gamma_{\rm cm}$ and found to be $\epsilon=C_d
(r_S/b)^{3d+3} \gamma_{\rm cm}^{2d+1}$, larger than previous
estimates  by many powers of $\gamma_{\rm cm}\gg 1$. This means that
in the ultrarelativistic case radiation loss becomes significant for
$b\gg r_S$, so radiation damping must be taken into account in
estimates of black hole production at  transplanckian energies. The
result is reliable for impact parameters in the overlap of
$\gamma^\nu r_S<b<b_c,\; \nu=1/2(d+1),\; {\rm and} \;b>\lambda_C$,
with $b_c$ marking (for $d\neq 0$) the loss of the notion of
classical trajectories and $\lambda_C\equiv \hbar/mc$ the Compton
length of the scattered particles.
\end{abstract}

\maketitle

Black hole (BH) production in LHC, predicted \cite{ADM}  \ in models
with TeV-scale gravity and large extra dimensions
\cite{ablt,ADD,GRW} about ten years ago, has been the subject of
intense theoretical study and numerical simulations (for a review
see \cite{reviews}). The prediction is based on the assumption that
for impact parameters of the order of the horizon radius
corresponding to the CM collision energy $2E=\sqrt{s}$
 \be r_S=\frac{1}{\sqrt{\pi}} \left[ \frac{8\Gamma
\left( \frac{d+3}{2}\right)}{d+2}\right]^{
\frac{1}{d+1}}\left(\frac{G_D
 \sqrt{s}}{c^4}\right)^{\frac1{d+1}}
\ee an event horizon should form due to the non-linear nature of gravity.
The $D=4+d$  dimensional gravitational constant is $G_D= {\hbar^{ d+1
}}/({M_*^{ d+2 } c^{d-1}})$, with $M_*$ the $D$-dimensional Planck mass. It is related to the
four-dimensional Planck mass $M_{\rm{Pl}}\equiv M_4$ via $
M_{\rm{Pl}}^2= M_*^{d+2}V, \;V=(2 \pi R)^d,$ where $R$ is the
large compactification radius.

This classical, essentially, picture of BH formation is
justified for transplanckian energies $ s\gg
G_D^{-2/(d+2)}=M_*^2$.  Indeed \cite{GiRaWeTrans}, in this case the
$D$-dimensional Planck length $
 l_{*}=\left( {\hbar G_D}/{ c^3}\right)^{ 1/({d+2})}=
  {\hbar}/{M_{*}c} $
 and the de Broglie length of the collision $
 \lambda_{B}=\hbar c/\sqrt{s}$
satisfy the {\it classicality} condition $\lambda_{B}\ll l_*\ll
r_S.$ Furthermore, gravity is believed to be the dominant force in
the transplanckian region. Thus, for BH masses large compared to
$M_*$, the use of classical Einstein theory is well justified.
Moreover, it seems that formation of BHs in four dimensions is
predicted by string theory \cite{Venez}. Thus, in spite of the fact
that there are issues which require further study \cite{cav}, a
consensus has been reached that the prediction of BHs in ultra-high
energy collisions is robust and is summarized in the widely accepted
four-stage process of formation and evaporation of BHs in colliders
\cite{ADM,YR}, namely (i) formation of a closed trapped surface
(CTS) in the collision of shock waves modeling the head-on particle
collision, (ii) the balding phase, during which the BH emits
gravitational waves and relaxes to the Myers-Perry BH, (iii) Hawking
evaporation and superradiance phase in which the experimental
signatures are supposed to be produced, and (iv) the quantum gravity
stage, where more fundamental theory like superstrings is important.
This scenario was implemented in computer codes \cite{compu} to
simulate the BH events in LHC, where they are expected to be
produced at a rate of several per second, and in ultra high energy
cosmic rays.

Here we focus on stage (i). Replacing the field of an
ultrarelativistic particle by that of a black hole in the infinite
momentum frame strictly speaking is only valid in the linearized
level, while the associated non-linear phenomena for particles may
be quite different from those in colliding waves. A typical
non-linear effect for particles is gravitational bremsstrahlung,
which is an important inelastic process in transplanckian
collisions. Apart from \cite{acv2}, where gravitational
bremsstrahlung of soft photons was studied in the context of string
theory, the existing estimates of gravitational radiation either
refer to phase (ii), or are based on the assumption of an already
existing BH (e.g. radiation from particles falling into the BH
\cite{car}), on results of linearized theory relevant only to the
case of non-gravitational scattering \cite{cav}, on weakly
relativistic numerical simulations \cite{Yo} or again on collisions
of waves in 4D \cite{DE}. For a related discussion see also
\cite{Mironov:2006wi}. However, a detailed study of gravitational
bremsstrahlung in the transplanckian regime in the ADD scenario was
missing. The purpose of this note is to present the results of such
a study (a more detailed account will follow \cite{GKST-4}). The
emitted energy is found to be larger than earlier estimates (see
e.g. \cite{GiRaWeTrans}) by a dimension dependent power of the
Lorentz factor of the collision. According to this result,
multidimensional gravitational radiation loss in transplanckian
collisions becomes significant already for impact parameters much larger
than the gravitational radius of the black hole. This means that
radiation reaction \cite{React} becomes essential and the picture of
colliding waves, which models particle motion with constant speed,
may not be adequate.

The standard ADD model assumes that empty space-time is the
product of the four-dimensional Minkowski space ${\mathcal M}_4$
(the brane) and a $d$-dimensional torus ${\mathcal T}^d$ and
treats gravity in the  linear approximation
$g_{MN}=\eta_{MN}+\varkappa_D h_{MN}$ ($\eta_{MN}$ has mostly
minuses). To calculate gravitational bremsstrahlung  classically
one has to extend the ADD setup beyond the linearized level. For
this, one expands the metric further as $h_{MN}\to h_{MN}+\delta
h_{MN}$ and adds the cubic interaction terms to the Fierz-Pauli
lagrangian. Equivalently, as in the standard theory of
gravitational radiation in four dimensions, one may expand the
$D$-dimensional Einstein tensor up to quadratic terms: \be
G_{MN}=-\frac{\varkappa_D}{2} \Box (\psi_{MN}+\delta \psi_{MN})-
\frac{1}{2} \varkappa_D^2 S_{MN}, \ee where the last term plays
the role of a gravitational stress-tensor, whose form is dimension
independent. Here $h\equiv \eta^{MN}h_{MN}$, $\psi_{MN}\equiv
h_{MN}-\eta_{MN} h/2$ is the trace-reversed metric perturbation,
$\Box\equiv \eta^{MN}\partial_M\partial_N$ is the $D$-dimensional
d'Alembertian and the coupling constant $\varkappa_D$ is defined
by $\varkappa_D^2\equiv16 \pi G_D$. Symmetrization and alternation
over indices is understood with $1/2$, while raising/lowering of
indices is meant with the flat metric.

Since $D$-dimensional Einstein equations imply the $D$-dimensional
Bianchi identities, which in turn imply the $D$-dimensional geodesic
equations for the particles, going beyond the linearized theory
might contradict the assumption of matter confinement on the brane.
However, it turns out that it is enough to assume that to zeroth
order in $\varkappa_D$ particles move on the brane. Then, the
corresponding zeroth order stress-tensor also lies on the brane
 \be T_{MN}(x^P)=\eta_M^\mu \eta_N^\nu T_{\mu\nu}(x)
\delta^d(\y).   \ee Its first order perturbation $\delta T_{MN}$,
due to the linearized gravitational interaction, is also confined
on the brane. Indeed, to first order the wave equation in the flat
harmonic gauge $\partial_M \psi^{MN}=0$ is   \be \lb{dal1}\Box
\psi_{MN}=-\varkappa_D T_{MN}.  \ee The source term is constructed
neglecting gravity, hence it is flat-space divergenceless too.
Since particles move freely in this order, this equation describes
non-radiative Lorentz-contracted gravitational potentials.
Radiation appears in second order in $\varkappa_D$ and is
described by the field $\delta \psi_{MN}$ satisfying (again in the
flat harmonic gauge) the equation   \be \lb{dal2}\Box \delta
\psi_{MN}=-\varkappa_D\tau_{MN},\quad \tau_{MN} = \delta
T_{MN}+S_{MN},  \ee where $S_{MN}$ is quadratic in the first order
gravitational potentials.  The right hand side of (\ref{dal2}) is
divergenceless by virtue of the expanded Einstein equations.
Therefore, the effective source of radiation is the sum of the
perturbation $\delta T_{MN}$ of the matter tensor, caused by the
first order gravitational interaction, and the gravitational stress
tensor $S_{MN}$, constructed from the first order metric
perturbations. This tensor is not confined on the brane, but
extends to the bulk. Note that $\delta T_{MN}$ and $S_{MN}$ are
not separately gauge invariant. Thus, beyond the linearized level
the effective source of radiation is not any more confined on the
brane.

Denote the $D$-dimensional coordinates as $x^M=(x^\mu,\, y^k)$,
where $x^\mu,\,\mu=0,1,2,3,$ lie on the brane and $y^k,\,
k=1,\ldots,d$ label the points of $T^d$. Imposing periodicity
conditions $h_{MN}(x,y^k+2\pi R)=h_{MN}(x,y^k)$ we obtain  an
infinite number of four-dimensional massive modes \be
h_{MN}(x^P)=\frac1{\sqrt{V }}\sum_{n\in \mathbb{Z}^d} h^{n}_{MN}(x)
\e^{i n_k y^k/R  }  \ee with masses $\mm^2=q_k^2,$ where $ q_k=
n_k/R$ is the quantized momentum on the torus, and which can be
regrouped into spin $2,1,0$ massive fields \cite{GRW,KT}. But, being
interested  in the total radiation in all modes, it is more
convenient to think about the metric perturbation as a
$D$-dimensional massless field with discrete momenta in the extra
dimensions. The four-dimensional fields $h^n_{MN}(x^\mu)$ are
further expanded into Fourier integrals defined by $\Psi(x)\!=\!\int
\Psi (q)\e^{-i q\cdot x} d^4 q/(2\pi)^4$.

Consider the small-angle collision of two point  masses $m,\, m'$.
Assuming that to zeroth order in $\varkappa_D$ their world-lines
lie on the brane $x^M=z^M(\tau),\,x'^M=z'^M(\tau')$, one verifies
that, as expected, they remain on the brane even after the
gravitational interaction is switched on. Therefore,
$z^M=z^\mu\delta^M_\mu$, \be \label{zz} z^\mu(\tau)=b^\mu+
\frac{p^\mu}{m}\tau+\delta z^\mu,\;\; z'^\mu(\tau)=
\frac{p'^\mu}{m'}\tau+\delta z'^\mu,   \ee where $p^\mu$ and
$p'^\mu$ are momentum parameters. Although the true initial
momenta $P^\mu\!=\displaystyle \!m\lim_{\tau\to-\infty}
\zt^\mu(\tau)$ and $ { P'}^\mu= \displaystyle
 m'\lim_{\tau\to-\infty} \zt'^\mu(\tau)$   differ from
$p^\mu,\,p'^\mu$,  they still satisfy $s=(P+P')^2=(p+p')^2$ \cite{elcl}. It is
convenient to work in the rest frame of
one of the particles, $m'$, choosing the coordinate axes on the
brane so that $ p'^\mu=m'(1,0,0,0),\; p^\mu=m\gamma(1,0,0,v),\;
 \gamma=1/\sqrt{1-v^2}$. Also, with no loss of generality one may
set $b^\mu=(0,b,0,0)$ and $b'^\mu=0$, so that $b$ is the impact
parameter. Finally, one may think  of brane localized vectors as
$D$-dimensional vectors with zero bulk components, e.g. $p^M=(p^\mu,
0,\ldots,0)$.

In the linearized theory the superposition principle implies that
both the zeroth and the first order energy-momentum tensor is the
sum of the contributions of the two masses, i.e. $T_{MN}=\,
\stackrel{m}{T}_{\!MN} +  \stackrel{\;m'}{T}_{\!\!MN}$ and $\delta
T_{MN}= \stackrel{\;\;\;m}{\delta\, T}_{ \!MN}  +
\stackrel{\;\;\;\;m'}{\delta\, T}_{ \!\!MN}$. The stress term
$S_{MN}$, on the other hand, represents a collective contribution,
which cannot be attributed to any one of the particles. Solving
the wave equation (\ref{dal2}), one finds that the radiation
amplitude consists of three terms, which correspond to radiation
from masses and from their gravitational stresses. This is similar
to the structure of the Born amplitude in quantum theory
\cite{GGM}, which in the case without extra dimensions and in the
low frequency limit is known to coincide with the classical
result. However, like in elastic scattering \cite{elcl} in ADD
$(d\neq 0)$, the classical treatment is essentially
non-perturbative in the quantum sense.

The spectral-angular distribution of the emitted energy in
gravitational bremsstrahlung during the collision is given by
 \be \lb{sp} \frac{dE_{\rm rad}}{d\omega
d\Omega}=\frac{G_D\omega^{2}}{2 \pi^{2} V}\sum_{ n \in \mathbb{Z}^d}
\sum_{\rm pol}   |\tau_{MN}(k) \varepsilon_{\rm pol}^{MN}|^2,
 \ee where \vspace{-.3cm} \be \lb{am}\tau_{MN}(k)=
\stackrel{\;\;\;m}{\delta\,
T}_{\!MN}(k)+\stackrel{\;\;\;\;m'}{\delta\, T}_{\!\!MN}(k)
+S_{MN}(k) \ee is the Fourier transform of the right hand side of
(\ref{dal2}). The first two terms have only the brane components
$M,N=0,1,2,3$, while the third is truly multidimensional. Introduce
next the {\em massive} four-dimensional wave-vector $k^\mu\equiv
(\sqrt{\omega^2+\mathrm{M}^2},\, \omega \tilde{\bf n}),$ with $
 \mathrm{M}^2={\bf \kappa}^2,\, \kappa_i=n_i/R$. The three-dimensional
unit vector $\tilde{\bf n}$ will be parameterized by the spherical
angles $\theta,\,\varphi$, in the usual way. Alternatively, one
can think of the radiation wave vector as the $D$-dimensional {\em
null} vector $k^M=(k^\mu,\, \kappa^i)$. The polarization tensors
of the emitted radiation are chosen according to the
$D$-dimensional picture. They are $D(D-3)/2$ symmetric transverse
traceless tensors orthogonal to $k^M$ and to each other. To
construct them start with $D-4$ unit space-like mutually
orthogonal vectors $e_{a}^M,\, a=3,4,\ldots,D-2$, orthogonal also
to $p^M,\,p'^M, \, k^M$ and $b^M$. Their contractions with
$\stackrel{\;\;\;m}{\delta\, T}_{\!MN}(k)$,
$\stackrel{\;\;\;\;m'}{\delta\, T}_{\!\!MN}(k)$ and $S_{MN}(k)$
will vanish. Take in addition the two vectors
\begin{align}
& e_1^M=\tilde{N}^{-1}\left[(k \cdot p) p'^M-(k\cdot p') p^M+\Bigl(
p\cdot p'-m'^2 \fr{k\cdot p}{k\cdot p'}\Bigr)k^M\right]
\nn \\
& e_2^M=\tilde{N}^{-1}\epsilon^{MM_1M_2M\ldots
M_{D-1}}p_{M_1}p'_{M2}k_{M_3}e_{3M_4}\ldots e_{D-2\;M_{D-1}},
\end{align}
(with $\tilde{N}^{2}=-\left[p (k \cdot  p') -p' (k \cdot  p)
\right]^2$) satisfying $e_2\cdot p =e_2\cdot p' =e_1\cdot p'=0.$
Using those, one builds two chiral graviton polarizations, which
 have direct four-dimensional analogs: \be
\varepsilon_{\pm}^{MN}=e_{\pm}^M e_{\pm}^N,\quad e_{\pm}^M=(e_1^M\pm
i e_2^M)/\sqrt{2}, \ee and a third one which also give non-zero
contribution in our case,  \be\label{3} \varepsilon_3^{MN}=
{\mathcal N}  \Big( \sum_a e_a^{M}e_a^{N} - ({D-4})
e_{+}^{(M}e_{-}^{N)} \!\Big),  \ee where ${\mathcal N}$ is a
normalization factor. The remaining polarization tensors contain at
least one vector from the set $\{e^M_a\}$ and they give zero being
contracted with $\tau_{MN}$.

The computation of the bremsstrahlung radiation proceeds as
follows. One finds the retarded fields generated by the
unperturbed particle trajectories and substitutes to the particle
geodesic equations to get the first order corrections to the
particles' motion. These are used to build the perturbations of
the particle energy-momentum tensors, whose Fourier transforms are
the first two terms in the radiation source (\ref{am}). The result
is expressed in terms of Macdonald (modified Bessel) functions with argument
$w_n=[w_0^2+\mm^2 b^2]^{1/2},\;w_0 =k\cdot p\, b/(m\gamma v)$ and
reads
\begin{align}
\label{dT} & \!\! \stackrel{\;\;\;m}{\delta\, T}_{\! \mu\nu}(k)=-
\frac{m m' \varkappa_D^2}{2\pi \gamma v^3 V } \; \e^{ik\cdot b}
\sum_{\n \in \mathbb{Z}^d}    \left[i \frac{\hat{K}_{1}
(w_n)}{(w_0)^2}\sigma_{\mu \nu}    +  2\frac{
p_{(\mu}\,p'_{\nu)}}{m m'} K_{0} (w_n) \gamma +
 \frac{p_{\mu}  p_{\nu} }{m^2} K_{0} (w_n)  \! \left( \! \gamma
\frac{w'_0}{w_0} \! - \! 1 \! \right) \!  \right],
\end{align}
where $w'_0\equiv k\cdot p'\,b/(m'\gamma v)$, $\sigma_{\mu\nu} = [ p_{\mu}
p_{\nu} k\cdot b-2k\cdot p\, p_{(\mu}b_{\nu) }]/m^2$ and
 $\hat{K}_{\lambda}(w) \equiv
 w^{\lambda}\,K_{\lambda}(w) $. The second term
$\stackrel{\;\;\;m'}{\delta\, T}{\!\!}_{\mu\nu}(k)$ is obtained
from (\ref{dT}) by the substitution $(m, m',  p_\mu, p'_\mu)\to
(m', m, p'_\mu, p_\mu)$ and $\e^{ik\cdot b}\to 1$.

 Next, one projects over polarizations. Only the first three contribute
in the chosen gauge, and all of them have zero contractions with the
$m'$ term in  (\ref{am}). So, the total amplitude receives two
contributions: (a) from the moving mass (\ref{dT}), and (b) from the
stress term in (\ref{am}), whose significance will be discussed
shortly.

The next step is to sum over the interaction modes. Assuming that a
large number of modes is excited, one can replace  summation over
$n^i$ by integration:
\begin{align}
\label{sum2int}
 \sum_{n} f_n \approx  \frac{V_d  \Omega_{d-1}}{(2\pi)^d}
\int f(\mathbf{q})\, q^{d-1}
 d q,
\end{align}
where $q^i=n^i/R$, where we have taken into account that  the
summand depends only  on  $  n_i^2$ . Performing integration we get
\cite{GKST-4}:
\begin{align}\label{sum2int1}
\sum_{ n \in \mathbb{Z}^d}\hat{K}_{\lambda}(w_n) \approx \left(
{2\pi R^2}/{b^2}\right)^{d/2}  \hat{K}_{\lambda+d/2}(w_0).
\end{align}
Summation over modes in the amplitude is the classical counterpart
of integration over transverse momenta of virtual gravitons, which
leads to tree-level divergences in ADD \cite{GRW}. However,
classically the result is finite like in elastic scattering
\cite{elcl}. The right hand side of \ref{sum2int1} is precisely the
same as we could obtain considering bremsstrahlung in an
uncompactified D-dimensional Minkowski space. Namely, the part of
the radiation amplitude corresponding to the fast particle   is
expressed in terms of the Macdonald functions of the argument $w_0$.
Properties of the Macdonald functions imply that the corresponding
radiation amplitude is concentrated in a narrow cone
$\theta<1/\gamma$ with a high frequency classical cutoff at
$\omega_c\sim 2\gamma^2/b$. This happens also in electromagnetic
bremsstrahlung in flat space.

However, in the case of gravitational interaction at hand the
spectral-angular distribution is very different, exactly because
of the special role of $S_{MN}$. Already in 4d gravity it was
shown that the contributions of (\ref{dT}) and $S_{\mu\nu}$ cancel
in the above spectral-angular region \cite{GGM}. Physically, this
is due to the fact that the radiation, emitted by particles
following ultrarelativistic time-like geodesics, follows null
geodesics, which are close to the former and give an effective
formation length of radiation in a given direction $\gamma$ times
larger than in flat space \cite{Khrip}. It is not surprising that
the situation is similar in ${\mathcal M}_4\times {\mathcal T}^d$
in the kinematical regime under discussion. The contribution from
the stresses $S_{MN}$, which is the classical counterpart of the
amplitude involving the three-graviton vertex, is rather
complicated. Nevertheless, it can be shown \cite{GKST-4} that like
for $d=0$ the leading term in the $S_{MN}$-amplitude for
$\gamma\gg 1$ exactly cancels (\ref{dT}). The next to leading
term, integrated over the transverse virtual momenta, is also
expressed in terms of Macdonald functions, with argument
$w'_0=\omega b/ \gamma $, which does not depend on angles.
Therefore, the radiation does not exhibit sharp anisotropy and the
frequency cutoff (from the condition $w'_0\sim 1$) is $\omega
\leqslant \omega_{\rm cr}= \gamma/ b .$

More thorough analysis shows that  only light emission modes  give
the leading contribution of the total emitted energy in the
ultrarelativistic collision.  For them the  projection of the total
amplitude over polarizations  $\varepsilon_\pm$ leads to (the third
polarization $\varepsilon_3$ gives a similar contribution):
\begin{align*}
  \tau_{\pm} \!= \!\frac{\varkappa_D^2 m m' (\gamma \xi)^{-1}}{ 4 (2 \pi)^{d/2+1}
  b^d}  \!  \left[ \!
(\cos 2\varphi \mp 2i)\cos \theta' \hat{K}_{d/2}(w'_0) \!+ \!
 \frac{\sin\theta' }{w'_0}  \! \left( \! (i\cos^3  \!\varphi\mp\sin 2\varphi)
  \hat{K}_{d/2+1}(w'_0)  \! - \! \sin 2\varphi (i  \sin \varphi \pm 1) \hat{K}_{d/2}(w'_0) \!
 \right) \!  \right] \!\!,
\end{align*}
where  $\xi=1-v \cos \theta$ and $\theta'$ is the radiation polar angle
in the rest frame of $m$ ($\sin\theta'=\sin\theta/\gamma \xi$).

Finally, upon integration of (\ref{sp}) over $\Omega$ and $\omega$
one obtains the radiated energy
\begin{align}
\label{gg6zr8} E_{\rm rad}= {\tilde C}_D \frac{m^2 m'^2
\varkappa_D^6 }{ b^{3d+3}} \gamma^{d+3}
\end{align}
with a known dimension-dependent coefficient. Qualitatively the
dependence on $b$ and $\gamma$ can be understood as follows. The
averaged over emission angles amplitude-squared is
$\langle|\tau_\pm|^2\rangle\sim b^{-2d}$, and $E_{\rm rad}\sim
\langle|\tau_\pm|^2\rangle\omega_{\rm cr}^3\, {\mathcal N}_{\rm eff}$,
where ${\mathcal N}_{\rm eff}\sim (R\gamma/b)^d$ is the effective number
of emitted light modes \cite{GKST-4}.

The expression (\ref{gg6zr8}) was obtained in the rest frame of the
particle $m'$. To pass to the CM frame, we calculate the relative
energy loss (radiation efficiency) $\epsilon\equiv E_{\rm rad}/E$,
and expresses the result in terms of the Lorentz factor in the CM
frame via (for $m=m'$) $\gamma_{\rm cm}^2=(1+\gamma)/2$: \be
\lb{loss} \epsilon=C_d \left(\frac{ r_S}{b}\right)^{3(d+1)}\!
\gamma_{\rm cm}^{2d+1}. \ee The two new features of (\ref{loss})
are: (a) the large factor $\gamma_{\rm cm}^{2d+1}$ due to the large
number of light KK modes involved both in the gravitational force
and in the radiation, and (b) the growing with $d$ coefficient:
\vspace{.3cm}
\begin{tabular}{|c||c|c|c|c|c|c|}\hline
  $\;d\; \displaystyle\vphantom{ \overrightarrow{d}} $ &   1 & 2 & 3 & 4 & 5&6  \\ \hline
   $\;C_d\; \displaystyle\vphantom{ \overrightarrow{d}} $ &\, 7.57 \,& 110 & \;\,1680\;\, & $2.6\times10^{4}$ & $4.1\times10^{5}$&$6\times10^{6}$ \\
   $\;r_S\;$ &   3.45 & 1.88 & 1.46 & 1.29& 1.21& 1.17 \\
   $\;b_c\;$ & 196 &  \;7.90 \;& 3.15& 2.11  & 1.72&1.53 \\
 \hline
 \end{tabular}
\vspace{.3cm}\\ The Table has $r_S$ and $b_c$ in TeV$^{-1}$
evaluated for $M_*\simeq 1\, {\rm TeV}$ and $\sqrt{s}\simeq 14 \,
{\rm TeV}$.  The classical description of small angle
ultrarelativistic scattering is strictly speaking valid for impact
parameter in the region $\gamma^\nu r_S\ll b\ll b_c,\;
\nu=1/2(d+1)$, where $b_c\equiv \pi^{-1/2} \left[ {\Gamma(d/2)G_D
s}/{\hbar c^5}\right]^{ {1}/{d}} \sim
r_S\left({r_S}/{\lambda_{B}}\right)^{{1}/{d}}$ is the scale beyond
which (for $d\neq 0$) the classical notion of particle trajectory is
lost \cite{GiRaWeTrans}. Another restriction comes from the quantum
bound on the radiation frequency $\hbar\omega_{\rm cr}< m\gamma$,
which is equivalent to $b> \lambda_C\equiv \hbar/(mc)$. For $d\neq
0$ the two conditions overlap provided $\lambda_C<b_c$. A rough
estimate of $\epsilon$ can be obtained in the case $\gamma^\nu
r_S\ll\lambda_C\ll b_c$ by setting $b=\lambda_C$ in (\ref{loss}),
which becomes $\epsilon=B_d (sm/M_*^3)^{d+2}$ ($ B_d=7.4, 0.6, 0.36,
0.45,
 0.81,  2.1,  7.0 $, for $d=0,1,\ldots,6$, respectively). Thus, a simple condition for strong
 radiation damping is
 \be
 sm\gtrsim M_*^3\;,
 \ee
 which may well hold for
heavy point particles with LHC energies. For example, in collisions
of particles with $m\sim {\mathcal O}(100 GeV)$ and energy
$\sqrt{s}\sim {\mathcal O}(10 TeV)$ all conditions for extreme
gravitational bremsstrahlung are satisfied for $d= 2$.

To conclude, a classical computation was presented of the
gravitational bremsstrahlung radiation in massive point-particle
collisions with transplanckian energies in 4+d dimensions. The
presence of the powers of $\gamma$ factors in our formula
(\ref{loss}), which, incidentally, agrees with the one obtained for
$d=0$ in \cite{TK}, implies enhanced and even extreme gravitational
bremsstrahlung and strong radiation damping in transplanckian
collisions. However, even though it is tempting and physically reasonable to do so, one
cannot strictly speaking apply (\ref{loss}) to the case of
proton-proton collisions in LHC. Protons are not point-particles,
while their constituents are too light and lie outside the region of
validity of our approximation. For the same reason formula
(\ref{loss}) cannot be applied to massless particle collisions,
which seem to require a different treatment \cite{acv2}.
Nevertheless, (a) one may have to include the reaction force
\cite{React} in the study of processes such as BH production, which
might even exclude the formation of a CTS. Note that there are
indications that gravitational collapse of an oscillating
macroscopic string does not take place, once gravitational radiation
is taken into account \cite{iengo}. Also, (b) bremsstrahlung, is a
strong process leading to missing energy signatures in
transplanckian collisions, which may further constrain the ADD
parameters.

Work supported in part by the EU grants INTERREG IIIA
(Greece-Cyprus) and FP7-REGPOT-2008-1-CreteHEPCosmo-228644. DG and
PS are grateful to the ITCP of the U. of Crete for its hospitality
in the early stages of this work. Their work was supported by the
RFBR project 08-02-01398-a. DG also thanks LAPTH Annecy-le-Vieux for
hospitality and support. We should thank S.~Giddings, A.~Mironov,
V.~Rychkov, M.~Sampaio and especially G.~Veneziano for useful
discussions.

\end{document}